\documentstyle[12pt]{article}

\begin{document}
\begin{center}
{\bf A Simple Mode on a Highly Excited Background:\\
Collective Strength and Damping in the Continuum}\\
V. V. Sokolov$^{(1)}$ and V. G. Zelevinsky$^{(1,2)}$\\
$^{(1)}$Budker Institute of Nuclear Physics,
630090 Novosibirsk, Russia\\
$^{(2)}$Department of Physics and Astronomy and\\
National Superconducting Cyclotron Laboratory,\\
Michigan State University, East Lansing, MI 48824-1321, USA
\end{center}

\begin{abstract}
Simple states, such as isobaric analog states or giant resonances, embedded
into continuum are typical for mesoscopic many-body quantum systems. Due to
the coupling to compound states in the same energy range, a simple mode
acquires a damping width (``internal" dynamics). When studied experimentally
with the aid of various reactions, such states reveal enhanced cross
sections in specific channels at corresponding resonance energies
(``external" dynamics which include direct decay of a simple mode and decays
of intrinsic compound states through their own channels).  We consider the
interplay between internal and external dynamics using a general formalism
of the effective nonhermitian hamiltonian and looking at the situation both
from ``inside" (strength functions and spreading widths) and from ``outside"
($S$-matrix, cross sections and delay times).  The restoration of isospin
purity and disappearance of the collective strength of giant resonances at
high excitation energy are discussed as important particular manifestations
of this complex interplay.
\end{abstract}
\newpage
\section{Introduction}

Dynamical features of open mesoscopic quantum
systems are characterized
by the presence of ``simple" (single-particle and collective) excitations,
``complicated" (chaotic) intrinsic motion involving many degrees of freedom,
and irreversible decay into continuum.
The coexistence and interplay of these phenomena
is the important aspect of all processes including excitation and deexcitation
of the system.
One of the questions of primary interest in nuclear physics,
especially for future experiments
with radioactive nuclear beams, is that of existence and purity of simple
modes of nuclear excitation embedded into continuum. Similar problems arise in
atomic and molecular physics,
physics of atomic clusters and mesoscopic solid state devices.

During the last decade, a number of related phenomena were discovered in this
area of nuclear physics, see for example \cite{Gaard,Snov}.
Saturation of the spreading width of the giant dipole
resonance (GDR) in hot nuclei \cite{sat,Enders,Brac95}, "disappearance" of the
collective strength of the GDR at high excitation energy  \cite{Gaard87}, and
existence and relatively narrow width of the double GDR
\cite{doubrit,doubsch,newd} are just few bright examples.
In the physics of isobaric analog states (IAS),
one can mention the evidence for existence of the so-called "broad poles"
\cite{Brent}, very weak fluctuations of the spreading widths of the IAS
throughout the periodic table \cite{iso,kuhl,comp},
and restoration of isospin purity at high
excitation energy \cite{Snov}.

In such problems, one always deals with a simple excitation
mixed with the dense background of complicated
states. The simple excitation is associated with a specific signal.
It can be a quantum
number which singles out the state in the ocean of surrounding states as it
happens in the IAS case. As a rule, such a state is
relatively pure with respect to this label when looked at
in the entrance channel.
The isospin purity is violated by the internal mixing \cite{iso}
when, due to the high
background level density, the statistical enhancement of perturbations
becomes extremely important, similar to the well known
enhancement of weak
interactions observed in parity nonconservation \cite{flam}.
The individuality of a "simple" mode can also be referred to its
specific structure, for example in the case of a
giant collective vibration, whose coherence
makes the state very different from the background. Such a
special state is characterized by a large multipole moment which
provides a strong collective gamma decay
\cite{Gaard}. In all cases, the manifestations of the simple mode in specific
reaction channels are intertangled with the chaotic mixing inside the system.

As a result of the mixing, the simple mode
is fragmented over exact stationary states which form the fine
structure of the spectrum.
Being averaged over the unresolved fine structure, the
excitation function is related to
the strength distribution of the original "label"
smoothly depending on excitation energy. More detailed
statistical analysis of observed
fluctuations, assuming generic correlations of energies and strengths for
the invisible underlying states,
is capable \cite{Darm} of extracting their characteristics.
In general, the strength functions and reaction cross sections represent two
sides of the process, internal and external, and the relation between them is
far from trivial. Thus, the strength distribution may or may not coincide with
the width distribution seen in the reactions and decays \cite{Alaga}.

The well known formalism \cite{BM} of the strength function
proceeds as if the states under consideration
were stable. However, all excited states, strictly
speaking, have a finite lifetime and therefore
belong to the continuum spectrum. The level widths of
the resonances in the continuum \cite{Fano,Weid}
are governed by the interaction which
is in general different from that forming
the discrete spectrum inside the system.
The effects of intrinsic mixing and coupling to and via
continuum have to be considered simultaneously.

Below we formulate a consistent quantum-mechanical approach which
fully accounts for the interplay of internal damping and decay and contains, as
particular cases, the ``disappearance" of the collective strength of the giant
resonance \cite{Gaard} explained by kinetic arguments in \cite{BBBB}, and the
restoration of isospin purity at high excitation energy \cite{Snov}
in accordance with the old idea by Morinaga \cite{Morin} and
Wilkinson \cite{Wilk}. We discuss the
general properties of the strength function
of a simple mode embedded into continuum in its relation to what is observed
in reactions.
Our consideration, being intentionally
schematic and less specific than in the well
known review paper \cite{iso}, is in many aspects complementary because of
its generality and
the simultaneous treatment of internal and external aspects of the problem.

\section{Effective hamiltonian}

We use the effective non-hermitian hamiltonian
\cite{Weid,Sok} in order to take into account internal
and external interactions on equal footing. The intrinsic structure at high
level density produces the set (``background") of
the basis intrinsic states $|n\rangle, \quad n = 1,
\dots N,$ where $N$ is supposed to be large. The simple state $|0\rangle$
is located in the the same range of energy.
All $N+1$
states have the same values of exact integrals of motion such as total
angular momentum. We assume that the basis states are characterized also
by quantities like isospin or parity which are approximate integrals of
motion. The
isospin mixing which is one of the subjects of our application is introduced
explicitly by the off-diagonal elements of the hamiltonian. Parity
nonconservation due to weak interactions can be another example of an
approximate conservation law which can be included in a similar manner.

The effective hamiltonian in $(N+1)$-dimensional space is the operator
\begin{equation}
{\cal H} = H - \frac{i}{2}W                      \label{1}
\end{equation}
containing two (real and symmetric
for a time reversal invariant system) matrices $H$ and $W$
which describe internal and external coupling, respectively.

The antihermitian part $W$ has a special structure \cite{Weid,iso,Sok}
being originated
by the on-shell decays into open channels $c=1,2,\dots,k$,
\begin{equation}
W = {\bf AA}^{T}\Rightarrow W_{nn'}=\sum_{c}A_{n}^{c}A_{n'}^{c}.   \label{2}
\end{equation}
Here we introduced the $(N+1)\times k$ matrices
${\bf A}= \{A_{n}^{c}\}$ of real transition amplitudes which are
proportional to the matrix elements of the full original
hermitian hamiltonian which connect intrinsic
and channel subspaces of total Hilbert space.

The hermitian part $H$ consists of the unperturbed
energy $\epsilon_{0}$ of the simple state $|0\rangle$,
the internal $N\times N$ hamiltonian $h$ describing the
background states $|n\rangle$, and the coupling between the simple
and complicated states. The real coupling matrix elements
$H_{0n}=H_{n0} \equiv V_{n}, n\geq 1,$ form an
$N$-dimensional column vector ${\bf V}$. The spectrum $h_{n}$
of the eigenvalues of $h$ is supposed to be very
dense. Along with the similarity of generic complicated wave functions
\cite{Horoi}, this justifies the statistical approach.

The effective hamiltonian ${\cal H}$ can be
studied with standard matrix methods \cite{Sok,Ann}. Its diagonalization gives
complex eigenvalues
\begin{equation}
{\cal E}_{j} = E_{j} - \frac{i}{2}\Gamma_{j}, \quad j=0,\dots,N, \label{3}
\end{equation}
and the quasistationary eigenstates $|j\rangle$ with a
pure exponential decay law $\sim \exp (-i{\cal E}_{j}t)$. The construction of
the effective hamiltonian guarantees the unitarity of the scattering matrix,
see below Sect. 5.2.

\section{Standard model of the strength function}

The description of the mixing of stable
internal states, which forms complicated
stationary superpositions and spreads the srength of original simple states,
is well known \cite{BM}.
With the antihermitian part $W$ omitted, the intrinsic propagation within
the closed system is described by
the Green function $G(E)$ of the hermitian part of the hamiltonian,
\begin{equation}
G(E) = \frac{1}{E - H}.                                 \label{7}
\end{equation}
The eigenvalues of the intrinsic
hamiltonian $H$ are given \cite{BM}
by the $(N+1)$ poles $E=\epsilon_{\alpha}$ of the Green function (\ref{7}).
 They are the roots of the secular equation
\begin{equation}
G_{00}^{-1}(E)\equiv
F(E)=E - \epsilon_{0} - \sum_{n\geq 1}\frac{V_{n}^{2}}{E-h_{n}}= 0.
                                                            \label{27}
\end{equation}

Each eigenfunction $|\alpha\rangle$ of $H$ carries a fraction
\begin{equation}
f^{\alpha} =|C^{\alpha}_{0}|^{2}= \Bigl(\frac{dF}{dE}
\Bigr)^{-1}_{E=\epsilon_{\alpha}}=
\left[1 + \sum_{n}V_{n}^{2}/(\epsilon_{\alpha} - h_{n})^{2}\right]^{-1}
                                                     \label{30}
\end{equation}
of the collective strength determined by the weight of the corresponding
component $C_{0}^{\alpha}$ in the expansion over the basis states,
\begin{equation}
|\alpha\rangle = C^{\alpha}_{0}|0\rangle + \sum_{n=1}^{N}C^{\alpha}_{n}
|n\rangle.                                                     \label{28}
\end{equation}
The smooth strength function of the simple excitation is defined in
terms of the average local level spacing $D$ of background states,
\begin{equation}
P_{0}(\epsilon) = [f^{\alpha}/D(\epsilon)]_{\epsilon_{\alpha}
= \epsilon}.                                                 \label{31}
\end{equation}
It is normalized according to $\sum_{\alpha}f^{\alpha}=
\int d\epsilon P_{0}(\epsilon) = 1$.

The formal solution (\ref{30}) requires the knowledge of statistical
properties of the background spectrum $h_{n}$ and
coupling matrix elements $V_{n}$.
The simplest ansatz used in the standard model \cite{BM} assumes
a roughly equidistant dense spectrum of $h_{n}$
and interaction intensities $V_{n}^{2}$ uncorrelated with energies $h_{n}$ and
slightly fluctuating around their mean value
$\langle V^{2} \rangle$. For convenience of the reader, we collect the results
of the uniform model in the Appendix, along with a brief discussion.

At $\langle V^{2}\rangle > D^{2}$,
the strength function of this uniform model has the Breit-Wigner shape
\begin{equation}
P_{0}(\epsilon) =
\frac{1}{2\pi}\frac{\Gamma^{\downarrow}}{(\epsilon - \epsilon_{0})^{2}
+ (\Gamma^{\downarrow})^{2}/4}.                            \label{32}
\end{equation}
with the spreading width given by the golden rule,
\begin{equation}
\Gamma^{\downarrow} \equiv \Gamma_{s} =
2\pi\frac{\langle V^{2}\rangle}{D}.    \label{33}
\end{equation}
The standard model just described is valid \cite{Bert,Lew,LBBZ}
if $\Gamma_{s}$ does not
exceed the energy range $\Delta E$ of coupling strength $V_{n}^{2}$
($\Delta E$ is defined by the spread of the doorway states which provide the
gates for the further mixing of the state $|0\rangle$).
This is expected to be a good approximation for the
IAS with the typical spreading width $\leq 100$ keV.
In the case of giant resonances
$\Gamma_{s}\simeq \Delta E $ and the uniform model
should be corrected \cite{Fraz}.
However the difference influences mainly the shape of the wings of the strength
function which is of minor importance for our purpose;
here we use the uniform model for definiteness.

\section{Simple state embedded into continuum}

\subsection{Formulation of the problem}

Now we take into account the openness of the system.
The simple state $|0\rangle$ is open to the direct decay (channels $c$
which display specific signatures of
the simple mode, for example collective $\gamma$-radiation from the giant
reasonance or pure isospin of the IAS).
Due to the intrinsic coupling to
compound states, the simple state also
acquires access to many ``evaporation channels" labeled by the superscript $e$;
partial widths depend on the distribution of strength of the simple mode
carried by specific compound states.

When applied to the IAS with isospin $T_{>}$, we have to consider
the surrounding background
states $|n\rangle$ which belong mainly to the isospin $T=T_{>}-1$. The
isospin mixing occurs mostly through intrinsic interaction \cite{iso}
so that the decay
channels for the decoupled simple mode and evaporation channels for compound
states carry different isospins. In many cases,
the effects we are interested in can be studied using one direct channel
which will be labeled as $c=0$.
Then we have in the hamiltonian (\ref{1}) the
amplitudes $A^{0}_{0} \equiv \sqrt{\gamma_{0}}$, where $\gamma_{0}$ is
the ``natural" width of the simple state, and $A^{e}_{n}, n\geq 1$.
All $A^{e}_{n}$ are assumed to be of the same order of magnitude.

At low energies (for example, for neutron resonances), only few decay channels
are open and the narrow compound states do not overlap. Their widths
$\gamma_{n}=\sum_{e}(A^{e}_{n})^{2}$ are small compared to their mean
energy spacing $D$. As energy and level density increase, we pass the
region of strong coupling via continuum where the width collectivization
occurs and broad "Dicke resonances" \cite{Rot,Sok,Ann}
form the contribution of direct processes. The situation changes again
when many uncorrelated decay channels are open, and the off-diagonal
elements (\ref{2}) of the antihermitian part $W$ of the effective
hamiltonian are averaged out.
Then the states $|0\rangle$ and $|n\rangle$ simply acquire finite
widths. These unstable states are coupled through the hermitian
interaction $V$ and this is what bridges the gap between the intrinsic
strength function and its manifestation in the resonance reactions.

To describe the open compound states, we introduce the $N\times N$ Green
function
\begin{equation}
g(z) = \frac{1}{z - h + (i/2)w}.                        \label{37}
\end{equation}
$N$ complex poles $z=\tilde{\epsilon}_{\nu}$
of $g(z)$ determine energies and evaporation widths of
compound resonances still decoupled from the simple mode.
In eq.(\ref{37}), $w$ stands
for the $(N\times N)$
submatrix of $W$, eq.(\ref{2}), which acts in the compound subspace and
describes the evaporation together with the interaction between
the compound states through common decay channels. The latter is
characterized by the off-diagonal matrix elements of $w$. Being the sums of
uncorrellated contributions of many evaporation channels,
$k\gg 1$, these elements,
due to mutual cancellations, are small in comparison with the diagonal
elements, $|w_{nn'}/w_{nn}|\sim 1/\sqrt k$ (see \cite{Sok}). The
corresponding corrections are of order of $\gamma^2_{ev}/kD^2$ where
$\gamma_{ev}$ is the typical evaporation width. We will neglect them below
assuming $\gamma_{ev}\ll \sqrt{k}D$. Under this condition, partial decay
widths of the compound states to specific evaporation channels are small,
$\gamma_{ev}/k\ll D$.

The complex energies of compound resonances in this approximation are equal
to $\tilde{\epsilon}_{\nu} = h_{\nu} - (i/2)\gamma_{ev}, \nu = 1,2,...,N$,
supposing on the statistical grounds that the fluctuations of widths of
compound states are weak since the number $k$ of evaporation channels is
large. The simple state has its own complex energy
$\tilde{\epsilon}_{0} = \epsilon_{0} - (i/2)\gamma_{0}$ where
$\gamma_{0}$ is the direct decay width.

Let us now switch on an interaction between the simple and compound
states through the hermitian coupling operator ${\bf V}$.
The mixing proceeds in
competition with the decays of intrinsic states, both via direct and
evaporation channels. Whence, we need to generalize
the standard procedure of determination of the strength function, Sect. 3,
for the decaying system. In our schematic although quite generic model, it
could be done exactly.

\subsection{Decay widths in the presence of intrinsic damping}

The diagonalization of the total non-hermitian hamiltonian
(\ref{1}) leads to $N+1$ complex eigenvalues (\ref{3}) which are
the roots $z = {\cal E}_{j}$ of the secular equation
(compare to (\ref{27}))
\begin{equation}
{\cal F}(z)\equiv z - \tilde{\epsilon}_{0} -
{\bf V}^{T}g(z){\bf V} = 0,                            \label{38}
\end{equation}
or, in the explicit form,
\begin{equation}
{\cal E}_{j}-\tilde{\epsilon}_{0}-\sum_{\nu}\frac{V_{\nu}^{2}}{{\cal E}_{j}
-\tilde{\epsilon}_{\nu}}=0.                              \label{38ex}
\end{equation}
The interaction amplitudes ${V}_{\nu}$, which couple the unstable simple state
$|{\sl 0}\rangle$ with complicated (and decaying as well) intrinsic states
$|\nu\rangle$, are still real in the approximation taken above (we neglected
the off-diagonal part of the continuum coupling $w$).

Similar to (\ref{28}), the quasistationary eigenstates $|j\rangle$ can be
represented as superpositions of decoupled unstable states
$|{\sl 0}\rangle,\dots , |\nu\rangle$
\begin{equation}
|j\rangle = \tilde{C}_{0}^{j}|{\sl 0}\rangle +
\sum_{\nu}\tilde{C}^{j}_{\nu}|\nu\rangle.                    \label{39}
\end{equation}
The fraction $\tilde{f}^{j} = |\tilde{C}_{0}^{j}|^{2}$ of the strength
of the simple state $|\sl 0\rangle$ carried by the quasistationary
state $|j\rangle$ is equal, as in eq.(\ref{30}), to
\begin{equation}
\tilde{f}^{j} =\frac{1}{1 + L^{j}},
\quad L^{j} ={\bf V}^{T}g^{\dagger}
({\cal E}_{j})g({\cal E}_{j}){\bf V}.               \label{40}
\end{equation}
With ${\cal E}_{j}=E_{j}-(i/2)\Gamma_{j}$, the loops $L^{j}$ can be written as
\begin{eqnarray}
L^{j} = \sum_{\nu}\frac{V_{\nu}^{2}}{|{\cal E}_{j} -
\tilde{\epsilon}_{\nu}|^{2}}
= \frac{2}{\Gamma_{j}-\gamma_{ev}}\;
{\rm Im}\sum_{\nu} V_{\nu}^{2}\;\frac{1}{{\cal E}_{j} -
\tilde{\epsilon}_{\nu}}.                               \label{41}
\end{eqnarray}
Using the secular equation (\ref{38ex}) we arrive at a very simple expression
\begin{equation}
L^{j} = \frac{\gamma_{0} - \Gamma_{j}}{\Gamma_{j} - \gamma_{ev}}, \label{42}
\end{equation}
leading to the individual strengths (\ref{40})
\begin{equation}
\tilde{f}^{j} = \frac{\Gamma_{j} - \gamma_{ev}}{\gamma_{0} - \gamma_{ev}}.
                                                                   \label{43}
\end{equation}

In other words, the resulting width of the quasistationary state $|j\rangle$
can be found from simple probabilistic arguments,
\begin{equation}
\Gamma_{j} = \gamma_{0}\tilde{f}^{j} +
\gamma_{ev}(1-\tilde{f}^{j}).                                   \label{44}
\end{equation}
The direct decay width is distributed over all quasistationary states
according to their fractions of the strength of the original simple state.
It is easy to check the normalization of the weights (\ref{43}):
\begin{equation}
\sum_{j}\tilde{f}^{j} = \frac{1}{\gamma_{0}-\gamma_{ev}}\Bigl(\sum_{j}
\Gamma_{j}-(N+1)\gamma_{ev}\Bigr) = 1                              \label{45}
\end{equation}
where the last step follows from the invariance of the imaginary part
of the trace of the hamiltonian (\ref{1}), $ \sum_{j}\Gamma_{j} = \gamma_{0}
+ N\gamma_{ev}$. We have to notice that the probabilistic interpretation
emerges here as a result of a strict quantum-mechanical calculation, with no
ensemble averaging or transition to a kinetic description.

\subsection{``Broad pole"}

Explicit expressions for the properties of the strength function, including
the spreading width along with
the decay widths into continuum can be obtained if
the average characteristics of the intrinsic spectrum and of
the coupling matrix elements are specified.

In the uniform model \cite{BM} used earlier for the stable states,
eq.(\ref{38ex}) gives a pair of coupled
equations for the real and imaginary parts of the complex energy (\ref{3}),
see Appendix,
\begin{equation}
E_{j} = \epsilon_{0} + \frac{1}{2}\Gamma_{s}\frac{x_{j}(1-y^{2}_{j})}
{1 + x^{2}_{j}y^{2}_{j}}, \quad
\Gamma_{j} = \gamma_{0} - \Gamma_{s}\frac{(1+x^{2}_{j})y_{j}}
{1 + x^{2}_{j}y^{2}_{j}}.                                \label{47}
\end{equation}
Here $\Gamma_{s}$ stands for the standard spreading width (\ref{33}), and
notations
\begin{equation}
x_{j} = \cot \left(\pi \frac{E_{j}}{D}\right),
\quad y_{j} = \tanh \left(\frac{\pi}{2}
\frac{\Gamma_{j} - \gamma_{ev}}{D}\right)                       \label{48}
\end{equation}
are introduced.

At moderate values of the interaction $V$, the simple state keeps
an appreciable fraction of the collective strength and preserves its
individuality, see eq. (\ref{57}) below. Such a state was called the broad
pole in \cite{Brent}. The problem of IAS can serve as a typical example.
The unperturbed analog state $|{\sl 0}\rangle$ arises at the energy
$\epsilon_{0}$ carrying almost pure isospin $T_{>}$. Its direct decay width
$\gamma_0$ is much larger than the evaporation width $\gamma_{ev}$ of
background states with isospin $T_{<}$ \cite{Weid}.  The isospin-violating
interaction $V$ mixes these states.

Assuming that the resulting width
$\Gamma_{0} \equiv \Gamma_{j=0}$ satisfies the condition
$(\Gamma_{0}-\gamma_{ev})\gg D$ we have from (\ref{48}) $y_{0}\approx 1$,
so that eqs. (\ref{47}) give for the complex root corresponding to the broad
pole
\begin{equation}
E_{0}\approx \epsilon_{0}, \quad \Gamma_{0} = \gamma_{0}-
{\Gamma}_{s}.                                      \label{56}
\end{equation}
The second expression reads $\Gamma = \Gamma^{\uparrow} - \Gamma^{\downarrow}$
in the notations chosen in \cite{Brent}. The state $T_{>}$ can be
observed only if it decays before mixing, $\gamma_{0}> \Gamma_{s}$.
The collective strength (\ref{43}) carried by the broad pole is then
\begin{equation}
\tilde{f}^{0} = \frac{\gamma_{0}- \gamma_{ev}- \Gamma_{s}}
{\gamma_{0} - \gamma_{ev}} =
1 - \frac{\Gamma_{s}}{\gamma_{0} - \gamma_{ev}}          \label{57}
\end{equation}
which remains of order of unity as long as $\Gamma_{0}$ noticeably exceeds
$\gamma_{ev}$. This formula extends to the case of unstable compound states
the measure introduced in \cite{Brent,McD} of the "purity of analog spin"
of the broad pole. On the other hand, the typical values $\tilde{f}^{j}$ for
$j\neq 0$ are small.

\subsection{General strength function}

The energy dependence of the strengths
(\ref{43}) is hidden in the secular equations (\ref{47}). Exclusion of
$x_{j}$ leads after simple algebra to the general equation for the
strength function which depends only on the
absolute value $|\gamma_{0}-\gamma_{ev}|$,
\begin{equation}
\tilde{f}^{j}=\frac{D}{2\pi |\gamma_{0}-\gamma_{ev}|}\;         \label{feq}
{\rm ln}\frac{(E_{j}-\epsilon_{0})^2+\frac{1}{4}\left[\Gamma_{s}+
|\gamma_{0}-\gamma_{ev}|(1-\tilde{f}^{j})\right]^2}{(E_{j}-\epsilon_{0})^2+
\frac{1}{4}\left[\Gamma_{s}-|\gamma_{0}-\gamma_{ev}|(1-
\tilde{f}^{j})\right]^2},
\end{equation}
or, for small $\tilde{f}_{j}$,
\begin{equation}
\tilde{f}^{j}=\frac{D}{2\pi |\gamma_{0}-\gamma_{ev}|}\;          \label{f}
{\rm ln}\frac{(E_{j}-\epsilon_{0})^2+\frac{1}{4}(\Gamma_{s}+
|\gamma_{0}-\gamma_{ev}|)^2}{(E_{j}-\epsilon_{0})^2+
\frac{1}{4}(\Gamma_{s}-|\gamma_{0}-\gamma_{ev}|)^2}.
\end{equation}

Substituting summation over $j$ by integration over
energy, one can easily check that this distribution is normalized as
\begin{equation}
\int \frac{dE_{j}}{D}\tilde{f}(E_{j})=\left\{
\begin{array}{cc} {\displaystyle
 \frac{\Gamma_{s}}{|\gamma_{0}-\gamma_{ev}|} }, & \ \ \ \ \  \Gamma_{s} <
|\gamma_{0}-\gamma_{ev}|;\\ {\ }\\
1, & \ \ \ \ \  \Gamma_{s} > |\gamma_{0}-\gamma_{ev}|. \label{N}
\end{array} \right.
\end{equation}
In the upper case, the contribution
$\tilde{f}^{0}=1-\Gamma_{s}/|\gamma_{0}-\gamma_{ev}|$
(compare with eq. (\ref{57})) of the simple state (\ref{56})
is lost in the integral. Indeed, the small factor in front of the logarithm
in eq. (\ref{feq}) is in this case compensated due to the small value of the
denominator of the expression under the logarithm so that eq. (\ref{f}) is not
valid for this special state. On the other hand, when the increasing mixing
rate characterized by the spreading width prevails upon the influence of
direct decays, the simple mode fully dissolves in the sea of compound states.

Except for an
exponentially narrow domain of parameters around the point
$\Gamma_{s}=|\gamma_{0}-\gamma_{ev}|$, the width (FWHM)
$\Gamma$ of the distribution
(\ref{f}) is determined by
\begin{equation}
\Gamma^{2}=|\Gamma_{s}^{2}-(\gamma_{0}-\gamma_{ev})^{2}|. \label{O}
\end{equation}
The tails of the strength function, $E\gg (\Gamma_{s}+|\gamma_{0}-
\gamma_{ev}|)$, are
universal and given by the standard
model, $\tilde{f}^{j}\approx (D/2\pi)\Gamma_{s}/E^{2}$.
In the limits $\Gamma_{s}\gg|\gamma_{o}-\gamma_{ev}|$ or
$\Gamma_{s}\ll|\gamma_{0}-\gamma_{ev}|$, eq.(\ref{f}) reduces to the
Breit--Wigner distribution
\begin{equation}
\tilde{f}^{j}=\frac{D}{2\pi}\left\{
\begin{array}{ll} {\displaystyle
\frac{\Gamma_{s}}{(E_{j}-\epsilon_{0})^2+\Gamma_{s}^2/4}
\leq \frac{2}{\pi}\;\frac{D}{\Gamma_{s}}, } \\ {\ } \\
{\displaystyle \frac{\Gamma_{s}}{|\gamma_{0}-\gamma_{ev}|}\;
\frac{|\gamma_{0}-\gamma_{ev}|}{(E_{j}-\epsilon_{0})^2+
|\gamma_{0}-\gamma_{ev}|^2/4}\leq \frac{2}{\pi}\;
\frac{D\Gamma_{s}}{(\gamma_{0}-\gamma_{ev})^2},   }        \label{Br-Wg}
\end{array}\right.
\end{equation}
respectively.
Near the point $\Gamma_{s}=|\gamma_{0}-\gamma_{ev}|$,
(\ref{f}) is invalid and eq.(\ref{feq}) gives
\begin{equation}
\tilde{f}^{0}=\frac{D}{\pi\Gamma_{s}}\left({\rm ln}\frac{2\pi\Gamma_{s}}{D}-
{\rm ln}\;{\rm ln}\frac{2\pi\Gamma_{s}}{D}+...\right).        \label{Ex}
\end{equation}
The strength $\tilde{f}^{0}$ is still larger than all
$\tilde{f}^{j}$ for $j\neq 0$ but this cannot influence the
normalization (\ref{N}). Fig. 1 illustrates the relation between the exact
expression for the strength function, eq.(\ref{feq}), the approximation
(\ref{f}), which is invalid in the center of the spectrum, and more crude
approximations (\ref{Br-Wg}).

The strength function gives an average
description of the fragmentation of individual simple configurations
in the intrinsic space. In the next section we study the problem as it is seen
in continuum properties.

\section{Scattering characteristics}

\subsection{Scattering matrix}

Up to now we concentrated on the ``inside" view of a simple unstable
mode mixed with complicated fine structure states. The ``outside" world was
present as a reservoir for irreversible decay through numerous open channels.
Now we take a glimpse of the same system from the viewpoint of reaction
amplitudes and cross sections where only asymptotic states
are observed.

The scattering matrix $\hat{S}=\{S^{cc'}\}$ at energy $E$ can be written as
\cite{Weid}
\begin{equation}
\hat{S}(E) = \hat{s}^{1/2}\{1 - i\hat{T}(E)\}\hat{s}^{1/2}, \label{4}
\end{equation}
\begin{equation}
\hat{T}(E) = {\bf A}^{T}{\cal G}(E){\bf A}.  \label{5}
\end{equation}
Here
$\hat{s}$ includes the potential scattering as well as channel coupling
and direct reactions in the continuum. Those effects
being unrelated to intrinsic dynamics are
irrelevant for our purpose, and $\hat{s}(E)$ can be considered
as a diagonal matrix with phase shift elements $\exp(2i\delta^{c})$
smoothly depending on $E$. The Green function in eq.(\ref{5}),
\begin{equation}
{\cal G}(z) = \frac{1}{z - {\cal H}},                         \label{6}
\end{equation}
describes the propagation governed by the total hamiltonian (\ref{1}).
It differs from the intrinsic Green function by the antihermitian
part of the effective hamiltonian.
Both ${\cal G}(z)$ and the scattering matrix (\ref{4})
have poles at the complex energies (\ref{3}).

It is a straightforward exercise to establish, with the aid of the
factorized structure (2) of the antihermitian part $W$, the relation
\begin{equation}
{\cal G}(E) = G(E) - \frac{i}{2}G(E){\bf A}\frac{1}{1 + (i/2)\hat{R}(E)}
{\bf A}^{T}G(E)                                           \label{8}
\end{equation}
between the two Green functions (\ref{6}) and (\ref{7}). The $\hat{R}$-matrix
in eq.(\ref{8}) is familiar from nuclear reaction theory \cite{Lane},
\begin{equation}
\hat{R}(E) = {\bf A}^{T}G(E){\bf A}.                      \label{9}
\end{equation}
It describes the propagation inside the closed system between two acts of
coupling to the continuum; the poles of $\hat{R}(E)$ correspond to the
energies $\epsilon_{\alpha}$ of intrinsic states with the mixing $V$ fully
accounted for.
The reaction matrix $\hat{T}(E)$ of eq.(\ref{5}) is similar to (\ref{9})
but includes all intermediate couplings to the continuum.
Finally, for the scattering matrix (\ref{4}, \ref{5})
the substitution (\ref{8}) gives
\begin{equation}
\hat{T}(E) = \frac{\hat{R}(E)}{1 + (i/2)\hat{R}(E)}, \quad
\hat{S}(E) = \hat{s}^{1/2}\frac{1 - (i/2)\hat{R}(E)}
{1 + (i/2)\hat{R}(E)}\hat{s}^{1/2}.                        \label{10}
\end{equation}

\subsection{Scattering wave function, delay time and unitarity}

The scattering wave function $|\Psi^{c}_{E}\rangle$ with the incident wave
in the channel $c$ at energy $E$ can be presented by the superposition
of intrinsic, $|n\rangle$, and continuum channel, $|c;E\rangle$, components,
\begin{equation}
|\Psi^{c}_{E}\rangle = \sum_{n}b_{n}^{c}(E)|n\rangle + \sum_{c'}
\int_{E^{c'}}^{\infty}dE'\chi^{cc'}(E,E')|c';E'\rangle    \label{12}
\end{equation}
where $E^{c'}$ is the threshold energy in the channel $c'$.
Recall that the decay amplitudes
$A_{n}^{c}$ are the matrix elements of the total original
hamiltonian between the states $|n\rangle$ and $|c;E\rangle$.
By a direct substitution of (\ref{12}) into the Schr\o dinger equation,
we find \cite{Weid} the $N\times k$ matrix ${\bf b}(E)$
of the intrinsic components $b_{n}^{c}$ as
\begin{equation}
{\bf b}(E) = {\cal G}(E){\bf A}\hat{s}^{1/2}.      \label{14}
\end{equation}

The diagonal elements of the $k\times k$ matrix
${\bf b}^{\dagger}(E){\bf b}(E)$ determine the norm of the internal part of
the wave function initiated in the channel $c$ at energy $E$. Therefore
this matrix should characterize the fraction of delay time in this
reaction due to intrinsic resonances. Indeed, the Smith's time delay
matrix is defined \cite{Smith} as
\begin{equation}
\hat{\tau}(E) = -i\hat{S}^{\dagger}(E)\frac{d\hat{S}(E)}{dE}.   \label{18}
\end{equation}
Taking into account only the resonance energy dependence
via the $\hat{R}$-matrix in eq.(\ref{10}), we find
\begin{equation}
\hat{\tau}_{res}(E) = -\hat{s}^{-1/2} \frac{1}{1-(i/2)\hat{R}(E)}
\frac{d\hat{R}(E)}{dE}\frac{1}{1+(i/2)\hat{R}(E)}\hat{s}^{1/2}.
                                                            \label{19}
\end{equation}
In the same resonance approximation one can neglect
the energy dependence of amplitudes ${\bf A}$ to get from (\ref{9})
\begin{equation}
(d\hat{R}/dE)_{res} = - {\bf A}^{T}G^{2}(E){\bf A}.    \label{19a}
\end{equation}
Using the relation (\ref{8}) between the Green functions $G$ and ${\cal G}$,
we obtain
\begin{equation}
\hat{\tau}_{res}=\hat{s}^{-1/2}{\bf A}^{T}{\cal G}^{\dagger}(E){\cal G}(E)
{\bf A}\hat{s}^{1/2}.                                    \label{21}
\end{equation}
Thus, the time
delay matrix (\ref{19}) coincides with the intrinsic norm matrix
found from (\ref{14}),
\begin{equation}
{\bf b}^{\dagger}{\bf b} = \hat{\tau}_{res}.       \label{20}
\end{equation}
The total Green function (\ref{6}) describes the propagation in the open
system and, therefore, the delay time as well.

We can now define the normalized probability $p_{n}^{c}(E)$ to find the
system in the intrinsic state $|n\rangle$ in the ``elastic" reaction
$c\rightarrow c$,
\begin{equation}
p_{n}^{c}(E) = \frac{1}{\tau^{cc}_{res}(E)}|b_{n}^{c}(E)|^{2}, \quad
\sum_{n}p_{n}^{c}(E) = 1.                                   \label{59}
\end{equation}
The probability $p_{0}^{c}(E)$ characterizes the weight of the simple state
$|0\rangle$ in the channel $c$. In the problem of the IAS
this quantity measures the isospin purity in a given channel.

The full scattering matrix (\ref{4},\ref{5}) is unitary provided the potential
scattering matrix $\hat{s}$ is unitary. It follows from the fact
that the decay amplitudes ${\bf A}$ in the entrance and exit channels
of eq.(\ref{5}) are the same which appear in all intermediate processes
described by the total propagator ${\cal G}(E)$ with the aid of the
effective hamiltonian (\ref{1},\ref{2}).

The unitarity condition $\hat{S}\hat{S}^{\dagger} = \hat{S}^{\dagger}
\hat{S} = 1$ gives for the reaction matrix (\ref{5})
\begin{equation}
\hat{T}^{\dagger}\hat{T} = i(\hat{T} - \hat{T}^{\dagger})      \label{15}
\end{equation}
which can be transformed, with the help of (\ref{2}) and (\ref{14}), into
\begin{equation}
\hat{s}^{1/2}{\bf b}^{\dagger}(E)W{\bf b}(E)\hat{s}^{-1/2} =
i\{\hat{T}(E) - \hat{T}^{\dagger}(E)\}.                        \label{16}
\end{equation}

\section{A simple case: Stable background states}

The simplest situation corresponds to the stable background states with
no direct access to open channels, $\gamma_{ev}\rightarrow 0$, when
the intrinsic evolution for the reaction in
the channel $c$ starts and ends at the simple state. The background states
are involved by the internal coupling only at the intermediate stages of the
reaction.
Calculating the diagonal element of the resonance time delay matrix
(\ref{21}) we obtain for the probability (\ref{59}),
\begin{equation}
p_{0}^{c}(E)\equiv f(E) =
\Bigl[1 + \sum_{n}\frac{V_{n}^{2}}{(E-h_{n})^{2}}\Bigr]^{-1}
 = [dF/dE]^{-1}.                       \label{69}
\end{equation}
This is nothing but the continuous generalization of the strengths
$f^{\alpha} = |C_{0}^{\alpha}|^{2}$ defined above by eq.(\ref{30}) in
discrete points $\epsilon_{\alpha}$ of the intrinsic energy spectrum,
$f^{\alpha} = f(E = \epsilon_{\alpha})$. Since the intrinsic states are
coupled to continuum through the state $|0\rangle$ and the probabilities
$p_{n}^{c}$ are normalized, eq.(\ref{59}), the decay (or population) partial
widths $\gamma_{0}^{c}$ do not appear in (\ref{69}). If several direct decay
channels $c$ are open, the energy behavior
(\ref{69}) is identical for all of them being determined by intrinsic
dynamics only.

The probability (\ref{69}) vanishes at energies $E=h_{n}$ of the
unperturbed background states which are located intermittently with the
actual energies $\epsilon_{\alpha}$. In the vicinity of $h_{n}$ the
complicated states dominate the intrinsic part of the scattering wave
function.

Another, though equivalent to (\ref{69}), representation of the
time delay in terms of the complex energies (\ref{3}) of quasistationary
states can be derived from (\ref{21}),
\begin{equation}
\tau^{cc}_{res}(E) = -2\frac{\gamma_{0}^{c}}{\gamma_{0}}
{\rm Im[ Tr} {\cal G}(E)]=\frac{\gamma_{0}^{c}}{\gamma_{0}}
\sum_{j}\frac{\Gamma_{j}}{(E-E_{j})^{2} + \Gamma_{j}^{2}/4}.
                                                        \label{75}
\end{equation}
The delay times for different channels $c$ are proportional to the
corresponding partial widths of the state $|{\sl 0}\rangle$ and have the
identical energy dependence determined by the complex energy
spectrum of intrinsic unstable states.

The representation (\ref{75})
is useful when the vicinity of the broad pole (\ref{56}) is considered.
It follows from (\ref{75}) that the contribution of this pole is a smooth
function of energy superimposed onto the picket fence of the $\delta$-like
peaks with the average value proportional to the Weisskopf recurrence time
$\pi/D$ for a long-lived wave packet. At the energy $E=E_{0}$, the time delay
in a channel $c$ due to excitation of the broad pole is equal to
$4\gamma_{0}^{c}/\gamma_{0}\Gamma_{0}$. On the other hand, one gets
\begin{equation}
|b_{0}^{c}(E_{0})|^{2}=\gamma_{0}^{c} |{\cal G}_{00}(E_{0})|^{2}\approx
4\gamma_{0}^{c}/(\gamma_{0})^{2}                         \label{75a}
\end{equation}
since the energy $E_{0}$ is very close to
the unperturbed energy of the state $|0\rangle$. Therefore the probability
maximum is determined by the fraction of the total width $\gamma_{0}$ of
the original mode which still resides at the broad pole,
\begin{equation}
p_{0}(E_{0}) = \frac{\Gamma_{0}}{\gamma_{0}}=
\frac{\gamma_{0} - \Gamma_{s}}{\gamma_{0}}. \label{76}
\end{equation}
in agreement with (\ref{57}) taken at $\gamma_{ev}=0$.

One should have in mind that the distribution
(\ref{69}) wildly fluctuates on the fine structure energy scale. With the
energy resolution worse than the level spacing $D$, one sees only a smooth
behavior coinciding with that of the strength function $P_{0}(E)$,
eq.(\ref{31}). It is quite
natural because here the intrinsic mixing is the only source for the
spreading of the strength, or for isospin impurity in the case of IAS.
An average magnitude of the probability to find the original isospin
can be easily estimated in the standard model with the uniform background.
Eq.(\ref{69}) gives here ($\Gamma^{\downarrow} = \Gamma_{s}$)
\begin{equation}
p_{0}(E) = \frac{\sin^{2}(E\pi/D)}{\sin^{2}(E\pi/D) + (\pi \Gamma^{\downarrow}
/2D)},                                                          \label{72}
\end{equation}
or, after averaging over fine structure, and taking $\Gamma^{\downarrow}
\gg D$,
\begin{equation}
\overline{p_{0}(E)} = \frac{1}{\pi}\frac{D}{\Gamma^{\downarrow}}. \label{73}
\end{equation}
This natural estimate (inverse number of fine structure states within the
spreading width) coincides with that used by von Brentano \cite{Brent}.

\section{Mixing with open compound states}
\subsection{Purity of a simple state}

The situation changes in the realistic case
with many open evaporation channels. Strong fluctuations
of the probability $p_{0}(E)$ are smeared out since the compound
poles are displaced
to the complex energy plane even with no coupling to the simple mode.
This probability remains considerable in
a finite vicinity of the point $\epsilon_{0}$ ensuring a noticeable isospin
purity of the internal part of the scattering wave function in this region.

If the simple mode and the compund states have no common decay channels,
the nonzero decay amplitudes are
$A_{0}^{c}=\sqrt{\gamma_{0}}$ (consider for simplicity a single direct decay
channel) and $A^{e}_{n}$. The reaction
amplitudes are equal to
\begin{equation}
T^{cc}(E)=\gamma_{0}{\cal G}_{00}(E), \quad
T^{ce}(E)=\sqrt{\gamma_{0}}
\sum_{\nu}{\cal G}_{0\nu}(E){A}^{e}_{\nu} \label{78}
\end{equation}
where now ${\cal G}^{-1}_{00}(E) = {\cal F}(E)$ (see (\ref{38})) whereas
\begin{equation}
{\cal G}_{0\nu}(E) = \frac{V_{\nu}}{E-\tilde{\epsilon}_{\nu}}{\cal G}_{00}(E).
                                                     \label{78*}
\end{equation}

The delay time in the elastic process, according to (\ref{21}) and
(\ref{14}), is given by
\begin{eqnarray}
\tau^{cc}_{res}(E)&=&\gamma_{0}|{\cal G}_{00}(E)|^{2}
\left(1+\sum_{\nu}\frac{V_{\nu}^{2}}{|E-\tilde{\epsilon}_{\nu}|^{2}}\right)
                                                \\   \nonumber
&\equiv&\gamma_{0}|{\cal G}_{00}(E)|^{2}\left[1+L(E)\right]        \label{79}
\end{eqnarray}
where the loop $L(E)$ is the analog of $L^{j}$, eq.(\ref{41}),
taken at the running real energy $E$ rather than at the complex energy
${\cal E}_{j}$. Therefore we find instead of (\ref{69})
\begin{equation}
p_{0}(E)\equiv\tilde{f}(E)=\frac{1}{1+L(E)}=\Bigl[1+\sum_{\nu}
\frac{V_{\nu}^2}{|E-\tilde\epsilon_{\nu}|^2}\Bigr]^{-1}.  \label{80}
\end{equation}
The function $\tilde{f}(E)$ extends the strength function (\ref{40}) of the
quasistationary states to a running real energy $E$ (compare with the similar
correspondence between the functions (\ref{30}) and (\ref{69}) in the case
of stable compound states). Note that, by definition (\ref{59}), the resonance
envelope $|{\cal G}_{00}|^{2}$ is divided out of normalized probabilities
$p_{0}(E)$ which behave uniformly within the spreading width.

The loop function (\ref{80}) can be calculated similar to (\ref{42}).
Under the same assumptions, it is equal to
\begin{equation}
L(E)=\frac{\Gamma_{s}}{\gamma_{ev}}y
\,\frac{(1+x^{2})}{1+x^{2}y^{2}}.                               \label{83}
\end{equation}
where, instead of (\ref{48}), we now have
\begin{equation}
x={\rm cot}\Bigl(\pi\frac{E}{D}\Bigr), \quad
y={\rm tanh}\Bigl(\frac{\pi}{2}\frac{\gamma_{ev}}{D}\Bigr).   \label{82}
\end{equation}

For a small evaporation width, $\gamma_{ev}\ll D$, the expression
(\ref{83}) reduces to
\begin{equation}
L(E)=\frac{\pi\Gamma_{s}}{2D}(1+x^{2}).                \label{84}
\end{equation}
The results in this limit do not depend on the evaporation width at all and
therefore coincide with those following from (\ref{69}). In particular, the
weight of the simple state in the intrinsic part of the scattering wave
function is in average of order of $D/\Gamma_{s}\ll 1$.

As level density and number of open channels increase, the ratio
$\gamma_{ev}/D$ rapidly grows together with the argument of $y$,
eq.(\ref{82}). One has a fast transition to the limit of the overlapping
background states when $y\approx 1$
and $L(E)\rightarrow \Gamma_{s}/\gamma_{ev}$. The
probability (\ref{80}) in this case is noticeably greater than in (\ref{73}),
\begin{equation}
\tilde{f} = \frac{\gamma_{ev}}{\gamma_{ev} +
\Gamma_{s}}\gg\frac{D}{\Gamma_{s}}.                          \label{n1}
\end{equation}
The fluctuations disappear, and the
simple state preserves its individuality in the intrinsic wave function
across the whole region of the giant or analog resonance. This behavior is
demonstrated in Fig. 2.

The purity of the
intrinsic part becomes perfect when $\gamma_{ev}\gg\Gamma_{s}$;
the depletion of admixed states of the opposite isospin occurs faster than
their population. This gives a
microscopic justification of the isospin purity at high excitation energy
predicted in \cite{Morin,Wilk} and recently observed experimentally \cite{Snov}.
At the same conditions, the fraction of the simple mode carried by a generic
compound state,
\begin{equation}
1-\tilde{f}=\frac{\Gamma_{s}}{\gamma_{ev}+\Gamma_{s}},       \label{n1a}
\end{equation}
is small. This means that the compound processes have no time to explore the
presence of the exceptional simple state.

The equivalent result was formulated
in terms of the kinetic balance between the processes of decay
and mixing in \cite{BBBB} where the mechanism for the
disappearance of the collective
strength of the GDR at high energies was suggested. The authors
showed that the probability of excitation of a collective mode in an
initially heated nucleus is equal, using our notations,
to $\Gamma_{s}/(\gamma_{ev} + \Gamma_{s})$
and therefore diminishes as $\Gamma_{s}/\gamma_{ev}$, when the
temperature exceeds a critical value determined by the condition
$\gamma_{ev}\sim\Gamma_{s}$. Complementary to the somewhat
qualitative kinetic arguments
of \cite{BBBB}, here the analogous conclusion
follows from a full quantum-mechanical consideration.

We need to mention parenthetically that such statements assume the
saturation of the spreading width $\Gamma^{\downarrow}\approx\Gamma_{s}$
as a function of
temperature. The absence of a considerable dependence on excitation energy
is well known for the IAS \cite{iso,kuhl,comp,PvB}. The saturation of the
intrinsic spreading width
presumably takes place for the GDR as well \cite{Gaard,sat,Enders,Brac95}.
General theoretical arguments in favor of such a saturation
\cite{iso,Lew,LBBZ,PvB,Fraz} are based on the
chaotization of the intrinsic dynamics and they will not be repeated here.

\subsection{Excitation and decay of a simple mode}

Here we compare the cross sections of various
processes initiated in the channel
$c$. They start with the excitation of the simple mode. The ``elastic"
scattering, $c\rightarrow c$, competes with the evaporation $c\rightarrow e$
through numerous compound channels $e$. These branches are described by the
amplitudes $T^{cc}$ and $T^{ec}$, respectively, see eq.(\ref{78}).

On the real energy axis, $z=E$, the uniform model leads to the inverse Green
function ${\cal G}_{00}^{-1}(E)={\cal F}(E)$, compare eq.(\ref{27}),
\begin{equation}
{\cal F}(E)= E-\epsilon_{0}-\Gamma_{s}\frac{x(1-y^{2})}{2(1+x^{2}y^{2})}
+\frac{i}{2}\Bigl[\gamma_{0}+
\Gamma_{s}y\frac{(1+x^{2})}{1+x^{2}y^{2}}\Bigr].      \label{81}
\end{equation}
At $y=0$ (no evaporation), the elastic cross section
\begin{equation}
 |T^{cc}|^2=
\gamma_{0}^2/|{\cal F}(E)|^2=\gamma_{0}^{2}
\left[\left(E-\epsilon_{0}-\frac{\Gamma_{s}}{2}\cot\frac{\pi E}{D}
\right)^{2} +\gamma_{0}^{2}/4\right]^{-1}          \label{81a}
\end{equation}
reveals fine structure fluctuations.
In the case of small $y\neq 0$, these fluctuations are enhanced
in a vicinity of the point $E=\epsilon_{0}$ due to the energy dependence of
the imaginary part of ${\cal F}(E)$, eq.(\ref{81}).
However, the fluctuations are washed
away when evaporation becomes strong, $\gamma_{ev}\gg D$, so that
$y\rightarrow 1$ and (\ref{81}) simplifies to
\begin{equation}
{\cal F}(E) = E-\epsilon_{0}+\frac{i}{2}\left(\gamma_{0}+
\Gamma_{s}\right).      \label{81*}
\end{equation}
Note that here the decay width $\gamma_{0}$ and the spreading width
$\Gamma_{s}$ are combined
into the total width of the resonance on the real energy axis. In Fig. 3
we illustrate the energy dependence of the elastic cross section $\sigma^{cc}
=|T^{cc}|^{2}$ for different values of relevant parameters.

Using the optical theorem, one obtains from eqs. (\ref{78},\ref{81})
and (\ref{83})
\begin{equation}
-2{\rm Im}\,T^{cc}(E) = |T^{cc}(E)|^2
\left[1 + \frac{\gamma_{ev}}{\gamma_{0}}L(E)\right]    \label{85*}
\end{equation}
for the total cross section initiated in the channel $c$.
The fraction of $|T^{cc}|^{2}$ in the total cross
section determines the branching ratio of
the simple decay mode,
\begin{equation}
{\cal B}^{cc}(E) = \frac{\gamma_{0}}{\gamma_{0} + \gamma_{ev}L(E)} =
\frac{\gamma_{0}\tilde{f}(E)}{\gamma_{0}\tilde{f}(E) +
\gamma_{ev}\left[1-\tilde{f}(E)\right]},                   \label{86}
\end{equation}
in agreement with the probabilistic interpretation of the function
$\tilde{f}(E)$. This function rather than its
discrete counterpart (\ref{40}) is relevant when an actual reaction process
is considered.

The amplitude $T^{ec}(E)$, eq.(\ref{78}), for evaporation in a given
channel $e$ after the simple state is excited in the entrance channel $c$,
strongly fluctuates together with the exit amplitudes $A^{e}_{\nu}$. This
amplitude vanishes in average.
Assuming many uncorrelated statistically equivalent decay channels, we can
use a natural statistical suggestion \cite{iso,Sok}
\begin{equation}
\langle A_{\mu}^{e}A_{\nu}^{e'}\rangle=\delta^{ee'}\delta_{\mu\nu}
\gamma_{ev}/k.                                              \label{86a}
\end{equation}
Taking into account eqs. (\ref{78}),(\ref{78*}), we obtain
\begin{equation}
\langle|T^{ec}(E)|^{2}\rangle
=\gamma_{0}|{\cal G}_{00}(E)|^{2}
\sum_{\mu\nu}\frac{V_{\mu}V_{\nu}\langle A^{e}_{\mu}A^{e}_{\nu}\rangle}
{(E-\tilde{\epsilon}^{\ast}_{\mu})(E-\tilde{\epsilon}_{\nu})}=
|T^{cc}|^2\frac{\gamma_{ev}}{k\gamma_{0}}L(E),                   \label{85}
\end{equation}
so that the corresponding branching ratio is equal to
\begin{equation}
{\cal B}^{ec} = \frac{1}{k}\,\frac{\gamma_{ev}L(E)}{\gamma_{0}
+\gamma_{ev}L(E)} =\frac{1}{k}\frac{\gamma_{ev}\left[1-\tilde{f}(E)\right]}
{\gamma_{0}\tilde{f}(E) +
\gamma_{ev}\left[1-\tilde{f}(E)\right]}.                   \label{86*}
\end{equation}
Eqs. (\ref{86}) and (\ref{86*}) give ${\cal B}^{cc}+k{\cal B}^{ec}=1$
in accordance with the unitarity condition. The statistical ansatz
(\ref{86a}) is self-consistent
because an equivalent approximation was in fact
introduced earlier when the off-diagonal elements of the antihermitian operator
$w$ in the compound space were substituted by the average evaporation width,
see the discussion after eq.(\ref{37}).

In the case of considerable evaporation and overlapping compound resonances,
$\gamma_{ev}/D\gg 1$, the branching ratios saturate at, see (\ref{n1}),
\begin{equation}
{\cal B}^{cc} = \frac{\gamma_{0}}{\gamma_{0}+\Gamma_{s}}, \qquad
{\cal B}^{ec} = \frac{1}{k}\,\frac{\Gamma_{s}}{\gamma_{0}+\Gamma_{s}}.
                                                                \label{B}
\end{equation}
For the saturated spreading width $\Gamma_{s}$, these limiting values
cease to be sensitive to the level density of compound states and depend on
excitation energy or temperature only through the direct width $\gamma_{0}$.
Under such conditions, only the simple state
with the total width $\gamma_{0}+\Gamma_{s}$, corresponding to the two
possible ways of its decay, escape and internal dissipation, is seen in the
scattering in the entrance channel $c$.
Here again the background of compound states serves as a reservoir for
irreversible decay, equivalent by its properties to decay into continuum.

\subsection{Reactions initiated in compound channels}

The processes started in the compound channels $e$, for example, driven by
a nuclear interaction of heavy ions, can populate the simple mode through
internal mixing. The corresponding amplitude $T^{ce}$ is the same as the
amplitude $T^{ec}$ considered above, eqs. (\ref{78}) and (\ref{85}). The
competing compound-compound processes are described by the set of the
amplitudes
\begin{equation}
T^{e'e}(E)=\sum_{\nu}\frac{A^{e'}_{\nu}A^{e}_{\nu}}{E-\tilde{\epsilon}_{\nu}}
+\sum_{\nu}\frac{A^{e'}_{\nu}V_{\nu}}{E-\tilde{\epsilon}_{\nu}}{\cal G}_{00}
\sum_{\mu}\frac{V_{\mu}A^{e}_{\mu}}{E-\tilde{\epsilon}_{\mu}}.     \label{la1}
\end{equation}
The second term in (\ref{la1}) accounts for the virtual excitation of the
simple mode with the subsequent deexcitation again via compound channels.

To evaluate the total cross section of compound-compound reactions,
\begin{equation}
\sigma^{e}\equiv \sum_{e'}\langle |T^{e'e}|^{2}\rangle,      \label{la1x}
\end{equation}
we perform
here the statistical averaging as in (\ref{86a}). Neglecting the numerical
corrections of the order $1/k$, and using the notations of Appendix
for the sums over the spectrum of the compound states, we obtain
\[\sigma^{e} = \frac{\gamma_{ev}^{2}}{k^{2}}
\{|{\cal S}|^{2}+k{\cal S}_{11}
+2{\rm Re}\left[{\cal G}_{00}^{\ast}\langle V^{2}\rangle
({\cal S}{\cal S}_{02}+k{\cal S}_{12})\right]\]
\begin{equation}
+ |{\cal G}_{00}|^{2}\langle V^{2}\rangle^{2}(|{\cal S}_{20}|^{2}
+k{\cal S}_{11}^{2})\}.                                \label{la2}
\end{equation}
The terms proportional to the number $k$ of open channels appear as a result of
pairwise coherent averaging of random decay amplitudes. Only these terms
survive in the limit $k\gg 1$. Taking the sums of Appendix in the overlapping
limit $y\rightarrow 1$ and recalling that in the same limit, according to
(\ref{85}),
\begin{equation}
\sigma^{ce}\equiv
|T^{ce}|^{2}=\frac{1}{k}|{\cal G}_{00}|^{2}\gamma_{0}\Gamma_{s}, \label{la3}
\end{equation}
we come, after many cancellations, to a simple expression for the total cross
section of all reactions initiated in a generic compound channel $e$,
\begin{eqnarray}
\sigma^{ce}+\sigma^{e}&\equiv&|T^{ce}|^{2}+\sum_{e'}\langle
|T^{e'e}|^{2}\rangle\\ \nonumber
&=& \frac{2\pi}{k}\left[(E-
\epsilon_{0})^{2}+\frac{1}{4}(\gamma_{0}+\Gamma_{s})^{2}\right]
|{\cal G}_{00}|^{2}\frac{\gamma_{ev}}{D}.                       \label{la4}
\end{eqnarray}
This gives the branching ratio for the deexcitation into the channel $c$
carrying the signature of the simple mode,
\begin{equation}
{\cal B}^{ce}=\frac{1}{2\pi}\,\frac{\gamma_{0}\Gamma_{s}}{(E-\epsilon_{0})^{2}
+(\gamma_{0}+\Gamma_{s})/4}\,\frac{D}{\gamma_{ev}}.          \label{la5}
\end{equation}
The resonance at the simple state is suppressed by the
inverse number $\gamma_{ev}/D$ of the background states on the typical
evaporation width. As it was discussed above, the observation of the
signal of the simple mode in the reaction started in a compound channel
becomes less probable with increasing $\gamma_{ev}$, in agreement with the
kinetic arguments of \cite{BBBB}.

The same result can be expressed with the aid of the
function $\tilde{f}$, eq.(\ref{80}),
\begin{equation}
\frac{\tilde{f}\sigma^{ce}}{(1-\tilde{f})\sigma^{e}}=
\frac{1}{2\pi}\gamma_{0}\Gamma_{s}|{\cal G}_{00}|^{2}
\frac{D}{\Gamma_{s}}.        \label{la6}
\end{equation}
Here $\tilde{f}\sigma^{ce}$ determines the fraction of the cross section of the
process $e\rightarrow c$ due to the intrinsic simple state; the denominator
is the similar fraction of the compound-compound cross section due to the
complicated intrinsic states, with no excitation of the simple mode.
The right hand side of (\ref{la6}) is the
resonance curve of the simple mode excited through the background (entrance
factor $\Gamma_{s}$) and deexcited through its own exit channel (factor
$\gamma_{0}$). The integral of the left hand side ratio over the energy region
covered by the spreading width gives the inverse number of fine structure
states in this region, $D/\Gamma_{s}$.

\subsection{Common decay channels}

One of the objections raised against the kinetic
explanation \cite{BBBB} of the disappearance of
the collective strength of the giant resonance is related to the possibility
of preequilibrium excitation of the giant mode \cite{Chom}. In this case the
intrinsic evolution would start with the state which already carries some
amount of collective strength. In our language such a possibility can be
taken into account via the presence of the reaction channels $a$ connected both
to the simple mode and to the background states.
For such channels, all amplitudes, $A^{a}_{0}$ and $A^{a}_{\nu}$ do not vanish;
until now we assumed that, before the internal mixing,
the simple state and the fine structure states have no common decay channels.
For the case of the IAS, this situation is associated with the extrnal
isospin mixing which is apparently of minor importance \cite{iso}. However,
for the giant resonance this effect can change the situation.

The common decay channels can be incorporated into theory without problems.
Here we consider the simplest case of a single common channel which can be
easily analyzed by the standard means.
The corresponding real amplitudes will be denoted
as $a_{0}$ and $a_{\nu}$ for the simple state and bacground states,
respectively. A many-channel case brings in many amplitudes of such type.
Being uncorrelated, they should not lead to any effects of coherent
enhancement.

In the single channel case, the exact algebraic solution gives the matrix
elements of the total Green function (\ref{6})
\begin{equation}
{\cal G}_{00}(E)=\left[E-\tilde{\epsilon}_{0}-\sum_{\nu}\frac{V_{\nu}^{2}}
{E-\tilde{\epsilon}_{\nu}}+\frac{i}{2}\frac{\alpha_{0}^{2}}{1+(i/2)R^{a}}
\right]^{-1},                                             \label{lb1}
\end{equation}
\begin{equation}
{\cal G}_{0\nu}(E)={\cal G}_{\nu 0}(E)={\cal G}_{00}\frac{u_{\nu}}{E-
\tilde{\epsilon}_{\nu}},                                  \label{lb2}
\end{equation}
\begin{equation}
{\cal G}_{\mu\nu}(E)=\frac{\delta_{\mu\nu}}{E-\tilde{\epsilon}_{\nu}}+
\frac{1}{E-\tilde{\epsilon}_{\mu}}\left(\frac{i}{2}\,\frac{a_{\mu}a_{\nu}}
{1+(i/2)R^{a}}+u_{\mu}{\cal G}_{00}u_{\nu}\right).              \label{lb3}
\end{equation}
Here the renormalized amplitudes are introduced for the decay of the simple
state through the channel $a$,
\begin{equation}
\alpha_{0}(E)=a_{0}+\sum_{\nu}\frac{V_{\nu}a_{\nu}}{E-\tilde{\epsilon}_{\nu}},
                                                             \label{lb4}
\end{equation}
and for the mixing between the simple state and the background including
the intermediate continuum states,
\begin{equation}
u_{\nu}(E)=V_{\nu}-\frac{i}{2}a_{\nu}\frac{\alpha_{0}}{1+(i/2)R^{a}}.
                                                               \label{lb5}
\end{equation}
The analog of the $R$-matrix, eq.(\ref{9}), for the $a$ channel is
\begin{equation}
R^{a}(E)=\sum_{\nu}\frac{a_{\nu}^{2}}{E-\tilde{\epsilon}_{\nu}}.  \label{lb6}
\end{equation}

Using these exact expressions we evaluate the reaction amplitudes. We are
interested in reactions starting in the channel $a$ and ending either in
the channel $c$ specific for our signature of the simple mode or in any of the
other channels, $a$ or $e$. The elastic $a\rightarrow a$ amplitude is given by
(compare (\ref{10})
\begin{equation}
T^{aa}= \frac{R^{a}}{1+(i/2)R^{a}}+\left(\frac{\alpha_{0}}{1+(i/2)R^{a}}
\right)^{2}{\cal G}_{00}.                                      \label{lb7}
\end{equation}
The deexcitation through the special channel $c$ is governed by the amplitude
\begin{equation}
T^{ca}=\sqrt{\gamma_{0}}({\cal G}_{00}a_{0}+\sum_{\nu}{\cal G}_{0\nu}a_{\nu})
=\sqrt{\gamma_{0}}{\cal G}_{00}\frac{\alpha_{0}}{1+(i/2)R^{a}}.    \label{lb8}
\end{equation}

To make a conclusion of the importance of the excitation through the common
channel, we assume that, similar to the amplitudes $A_{\nu}^{e}$,
eq.(\ref{86a}), the new amplitudes $a_{\nu}$ are uncorrelated quantities with a
large magnitude which contributes significantly to the total width
$\gamma_{ev}$ of the compound states, $\langle a^{2}\rangle
\equiv \gamma_{a}\sim \gamma_{ev}
\gg D$. Using the estimates of the Appendix for the sums over the fine
structure states in the overlapping limit, we obtain $R^{a}\approx-i\pi(
\gamma_{a}/D)$. Thus, $R^{a}$ is a large imaginary quantity determined by the
number of compound states in the interval $\gamma_{a}$. According to the same
estimates, the first term in (\ref{lb7}) dominates, and the contribution of
terms containing the sum with the cross products $V_{\nu}a_{\nu}$ is relatively
small. Finally,
\begin{equation}
\frac{\langle |T^{ca}|^{2}\rangle}{\langle |T^{aa}|^{2}\rangle}\approx
\left(\frac{D}{\pi \gamma_{a}}\right)^{2}\gamma_{0}\gamma_{0}^{a}
|{\cal G}_{00}|^{2}.                                            \label{lb9}
\end{equation}
The average partial width for the decay of the simple state into the channel
$a$ is equal to
\begin{equation}
\gamma_{0}^{a}\equiv\langle |\alpha_{0}|^{2}\rangle=a_{0}^{2}+
\frac{\gamma_{a}}{\gamma_{ev}}\Gamma_{s}.                   \label{lb10}
\end{equation}
This, quite general, result shows that, in the case of the common channel
capable of populating both, simple and compound states,
the statistical
branching for the deexcitation via the simple mode drops with the
increasing level density $\rho\sim 1/D$
of compound states. Whence, strong common channels
with $\gamma_{a}\sim\gamma_{ev}$ cannot recover the disappearing simple mode.
Weak channels, $\gamma_{a}\simeq \gamma_{ev}/k$,
are useless because of their small total cross sections.

\section{Conclusion}

In the paper we considered the most general properties of an open quantum
system where a simple mode of excitation interacts with the background of very
complicated states. Both, simple and compound, states are coupled to the
continuum and have finite lifetimes. Internal dynamics (mixing) and external
dynamics (decays) are intertangled in a nontrivial way. The intrinsic dynamics
in the presence of the continuum
are governed by the effective nonhermitian hamiltonian. The widths of intrinsic
states modify the strength function of the simple mode. This view from
``inside" has to be supplemented by that from ``outside"
for determining the observables measured in a real scattering experiment,
such as cross sections and delay times.
In a formal language, here we project the dynamics of quasistationary intrinsic
states back to the real energy axis. The effective hamiltonian by its
construction guarantees the correct properties
of the scattering matrix including unitarity. Therefore it becomes possible to
use the knowledge of internal dynamics in order to compare cross sections of
competing processes.

The general although schematic character of the analysis allows one to draw the
conclusions concerning the manifestations of the simple mode in various
situations. A typical example is given by the IAS which can be seen as a broad
pole \cite{Brent} or to be dissolved in the sea of the fine structure levels of
another isospin.
The analysis, analogous to that in \cite{iso},
confirms the old idea \cite{Morin,Wilk} of increasing
isospin purity of the IAS at high excitation energy. The experimental data
\cite{Snov} agree with this conclusion. The isospin purity is restored because
of the very fast depopulation of the admixed background states when their
decay width $\gamma_{ev}$ increases compared to the spreading width
$\Gamma_{s}$ of the simple state (IAS in this case). Here $\Gamma_{s}$ is
assumed to be a slow changing or saturating function of excitation energy
\cite{Brac95} as
predicted by the analysis based on the chaotic character of the intrinsic
dynamics \cite{LBBZ}.

Such a consideration is not specific for the IAS and can be applied to other
simple modes embedded into continuum. The giant dipole resonance is known to
preserve its individuality up to high excitation energy or temperature
\cite{Gaard}. In particular, this is clearly seen in the observation of the
nearly harmonic double-phonon excitations \cite{newd}.
The new phenomenon of disappearance of the collective strength of the GDR
\cite{Gaard} is still a debatable subject. Such a behavior was qualitatively
explained in \cite{BBBB} as a result of a shift of the kinetic equilibrium in
favor of compound decays when the ratio $\gamma_{ev}/\Gamma_{s}$ increases.
Our general quantum-mechanical analysis confirms this result. Moreover, we
made the arguments which demonstrate that the conclusion is still valid when
the simple mode can be excited from the reaction channels which are common
for the simple mode and the background states.

To complete the analysis, it would be interesting to consider the situation
when several simple states can share the collective strength and the decay
width into the channel which signals the deexcitation of the simple mode. This
is the case in the realistic calculation of the GDR. The collective peak
accumulating a large part of the isovector dipole strength is shifted to high
energies compared to the unperturbed shell model position. However, some
strength is still concentrated at the unshifted energy. This ``configuration
splitting" leads to specific interference phenomena \cite{Alaga} which again
can be described with the use of the effective nonhermitian hamiltonian.
The distribution of the dipole strength and the width evolves with the
increasing excitation energy which should be taken into account when the
interplay of the internal interaction and external decays is considered.
Typically, this results in the quenching of the collective strength and its
redistribution in favor of the low energy component. These effects \cite{Phil}
are discussed in \cite{SRSM}.\\[0.5cm]

The authors are indebted to P. von Brentano who initiated this work and made
an important impact by numerous discussions at the initial stage.  We thank
D.V.Savin for constructive discussions and assistance.  One of us (V.S.) is
grateful to Y.Fyodorov, F.Izrailev, I.Rotter and H.-J.Sommers for
interesting discussions. He also thanks the K\"{o}ln University and the
National Superconducting Cyclotron Laboratory for generous hospitality.
This work was supported by the National Science Foundation, through grants
94-03666 and 95-12831, and by the INTAS grant 94-2058.

\begin{center}
{\bf Appendix}\\[0.3cm]
A uniform model of compound spectra
\end{center}

Assuming the equidistant spectrum of unstable
background states with the level spacing $D$ and the decay width $\gamma$,
which corresponds to $\gamma_{ev}$ of the main text, and substituting actual
coupling matrix elements $V_{\nu}^{2}$ by their average $\langle V^{2}\rangle$,
we have to deal with the sums as the trace of the Green function (\ref{37})
\begin{equation}
{\cal S}=\sum_{\nu=-\infty}^{\infty}\zeta_{\nu}(E).             \label{A1}
\end{equation}
For the calculations of the scattering processes, the energy $E$ is real and
\begin{equation}
\zeta_{\nu}(E)=\frac{1}{E-\tilde{\epsilon_{\nu}}}=\frac{1}{E-\nu D+(i/2)\gamma}.
                                                                    \label{A2}
\end{equation}
The summation in (\ref{A1}) leads to
\begin{equation}
{\cal S}=\frac{\pi}{D}\cot\left[\frac{\pi}{D}\left(E+\frac{i}{2}\gamma
\right)\right]=\frac{\pi}{D}\,\frac{x-iy}{1+ixy}      \label{A3}
\end{equation}
where the parameters are introduced
\begin{equation}
x=\cot\left(\frac{\pi E}{D}\right), \quad y=\tanh\left(\frac{\pi\gamma}{2D}
\right).                                                     \label{A4}
\end{equation}

As the decay width $\gamma$ increases, the quantity $y$ changes very rapidly
from a small value $y\approx \pi\gamma/2D$ for isolated long-lived states,
when $\gamma/D\ll 1$, to a value exponentially close to 1 for overlapping
levels, when $\gamma/D\gg 1$. In practice it is sufficient to consider just
these two limiting cases. At small $\gamma$, the imaginary part of ${\cal S}$
is small, $\propto y\approx(\pi \gamma/2D)$,
and the real part of ${\cal S}$ is equal to
$\pi x/D$ as for stable levels \cite{BM}. In the opposite case of large
$\gamma/D$, the real part vanishes $\sim(1-y^{2})$ whereas Im${\cal S}\approx
-\pi/D$. Both cases have a general meaning being not limited by restrictions of
the uniform model. Thus, the result for the overlapping
case follows immediately
after substituting the $\sum(E-\epsilon_{\nu})^{-1}$ by the integral over the
levels with a level density $1/D$ and using a small shift of energies into the
complex plane. This expression is routinely used in statistical theory of
nuclear reactions \cite{iso}. A similar consideration is valid for the sums
as in (\ref{38ex}) and (\ref{48})
taken at a fixed complex energy ${\cal E}_{j}=E-(i/2)\Gamma_{j}$ instead
of running real energy $E$.

More complex sums can be easily analyzed in the same way. Here we give some
examples used in the text (the notation ${\cal S}_{mn}$ corresponds to $m$
factors $\zeta_{\nu}$ and $n$ factors $\zeta_{\nu}^{\ast}$ so that the basic
sum ${\cal S}\equiv {\cal S}_{10}$):
\begin{equation}
{\cal S}_{20}=\sum_{\nu}\zeta_{\nu}^{2}=\left(\frac{\pi}{D}\right)^{2}
\frac{(1-y^{2})(1+x^{2})}{(1+ixy)^{2}},                   \label{A5}
\end{equation}
\begin{equation}
{\cal S}_{11}=\sum_{\nu}|\zeta_{\nu}|^{2}=\frac{2\pi}{D\gamma}\,
\frac{y(1+x^{2})}{1+x^{2}y^{2}},                                 \label{A6}
\end{equation}
\begin{equation}
{\cal S}_{12}=\sum_{\nu}|\zeta_{\nu}|^{2}\zeta_{\nu}^{\ast}
=\frac{i\pi}{D\gamma}\,
\frac{1+x^{2}}{1+x^{2}y^{2}}\left[\frac{2y}{\gamma}-\frac{\pi}{D}(1-y^{2})
(1-x^{2}y^{2}+2ixy)\right].                                  \label{A7}
\end{equation}
In the overlapping limit $y\rightarrow 1$, these
sums go to 0, $2\pi/D\gamma$ and
$2i\pi/D\gamma^{2}$, respectively. The first sum (\ref{A5})
vanishes in accordance with
the fact that both poles in the equivalent integral are located on the same
side of the real axis. The nonvanishing sums are proportional to the level
density $\rho=1/D$, i.e. they have a coherent component growing at high
excitation energy.

The sum ${\cal S}_{22}$ can be calculated as
\begin{equation}
{\cal S}_{22}=\sum_{\nu}|\zeta_{\nu}|^{4}=\left[\frac{\partial^{2}}
{\partial E \partial E'}\sum_{\nu}\zeta_{\nu}(E)\zeta_{\nu}^{\ast}(E')
\right]_{E'=E}.                                           \label{A8}
\end{equation}
After simple algebra, we obtain
\begin{equation}
{\cal S}_{22}=-\frac{4}{\gamma^{3}}{\rm Im}{\cal S}+\frac{2}{\gamma^{2}}
{\rm Re}\frac{d{\cal S}}{dE}.                            \label{A9}
\end{equation}
In the overlapping limit, the first term in (\ref{A9}) gives $4\pi/D\gamma^{3}$
whereas the second one is proportional to $(1-y^{2})/D^{2}\gamma^{2}$ and
therefore it is small compared to the first term
since the exponential smallness of $(1-y^{2})$ overcompensates
an extra factor $\gamma/D$.

\newpage

{\bf Figure captions}\\[0.2cm]

{\bf Figure 1}. The strength function $\tilde{f}^{j}$ as a function of energy
for the values of parameters $\Gamma_{s}/D=100$ and $|\gamma_{0}-\gamma_{ev}|
/D=90$. The solid curve gives the exact numerical solution of eq.(\ref{feq}),
the dotted line corresponds to the approximation (\ref{f}), the dash-dotted
curves show the Breit-Wigner approximations (\ref{Br-Wg}). \\[0.3cm]

{\bf Figure 2}. The relative probability $\tilde{f}(E)$, see eqs. (56-58),
of excitation of a simple state through the channel $c$ as a function of
energy, $E/D$, and the evaporation width, $\gamma_{ev}/D$. The value of the
spreading width, $\Gamma_{s}/D=10$ is chosen for illustrative purposes to make
the oscillations along the energy axis clearly seen; the oscillations rapidly
disappear as $\gamma_{ev}$ grows.\\[0.3cm]

{\bf Figure 3}. Elastic cross section $\sigma^{cc}$ in the channel $c$ as a
function of energy. The parameters in part {\sl (a)} are $\gamma_{0}/D=10$ and
$\Gamma_{s}/D=5$; the cross section is shown for different evaporation widths,
$\gamma_{ev}/D$, which correspond to the values $y=0.3$ (dots), 0.7 (solid
curve) and 1 (dash-dotted curve). The situation with $\Gamma_{s}>\gamma_{0}$
is shown, for the same values of $y$, in part {\sl (b)} where $\gamma_{0}/D
=5$ and $\Gamma_{s}/D=10$; note the different scale for the cross section.
Part {\sl (c)} shows the cross section
for $\gamma_{0}/D=100$ and $\Gamma_{s}/D=15$ with $y=0.5$ (oscillatory curve)
and $y=1$ (thick curve).


\newpage
\begin{thebibliography}{99}
\bibitem{Gaard}J.J.Gaardh\o je. Ann. Rev. Nucl. Part. Sci. {\bf 42} (1992) 483.
\bibitem{Snov}K.A.Snover. Nucl. Phys. {\bf A553} (1993) 153c.
\bibitem{sat}A.Bracco {\sl et al}. Phys. Rev. Lett. {\bf 62} (1989) 2080.
\bibitem{Enders}G.Enders {\sl et al}. Phys. Rev. Lett. {\bf 69} (1992) 249.
\bibitem{Brac95}A.Bracco {\sl et al}. Phys. Rev. Lett. {\bf 74} (1995) 3748.
\bibitem{Gaard87}J.J.Gaardh\o je {\sl et al}. Phys. Rev. Lett. {\bf 59}
(1987) 1409.
\bibitem{doubrit}J.Ritman {\sl et al}. Phys. Rev. Lett. {\bf 70} (1993) 533.
\bibitem{doubsch}R.Schmidt {\sl et al}. Phys. Rev. Lett. {\bf 70} (1993) 1767.
\bibitem{newd}K.Boretzky {\sl et al}. Phys. Lett. {\bf 384} (1996) 30.
\bibitem{Brent} P. von Brentano. Z. Phys. {\bf A306} (1982) 63.
\bibitem{iso}H.L.Harney, A.Richter and H.A.Weidenm\"{u}eller. Rev. Mod. Phys.
{\bf 58} (1986) 607.
\bibitem{kuhl}E.Kuhlmann. Z. Phys. {\bf A322} (1985) 527.
\bibitem{comp}J.Reiter and H.L.Harney. Z. Phys. {\bf A337} (1990) 121.
\bibitem{flam}V.V.Flambaum. Phys. Scr. {\bf T46} (1993) 198.
\bibitem{Darm}G.Kilgus {\sl et al.} Z. Phys. {\bf A326} (1987) 41.
\bibitem{Alaga}V.V.Sokolov and V.G.Zelevinsky. Fizika {\bf 22} (1990) 303.
\bibitem{BM}A.Bohr and B.Mottelson, {\sl Nuclear Structure}, vol. I,
Benjamin, New York, 1969.
\bibitem{Fano}U.Fano. Phys. Rev. {\bf 124} (1961) 1866.
\bibitem{Weid} C.Mahaux and H.A.Weidenm\"{u}ller. {\sl Shell-Model
Approach to Nuclear Reactions}, North-Holland, Amsterdam, 1969.
\bibitem{BBBB}P.F.Bortignon, A.Bracco, D.Brink, and R.A.Broglia.
Phys. Rev. Lett. {\bf 67} (1991) 3360.
\bibitem{Morin}H.Morinaga. Phys. Rev. {\bf 97} (1955) 444.
\bibitem{Wilk}D.H.Wilkinson. Philos. Mag. {\bf 1} (1956) 379.
\bibitem{Rot}P.Kleinw\"{a}chter and I.Rotter. Phys. Rev. {\bf C32} (1985) 1742.
\bibitem{Sok}V.V.Sokolov and V.G.Zelevinsky. Nucl. Phys. {\bf A504}
(1989) 562.
\bibitem{Horoi} V.Zelevinsky, M.Horoi and B.A.Brown. Phys. Lett. {\bf B350}
(1995) 141; M.Horoi, B.A.Brown and V.G.Zelevinsky. Phys. Rev. Lett. {\bf 74}
(1995) 5194.
\bibitem{Ann}V.V.Sokolov and V.G.Zelevinsky. Ann. Phys. {\bf 216} (1992) 323.
\bibitem{Bert}C.A.Bertulani and V.Zelevinsky. Nucl. Phys. {\bf A568}
(1994) 931.
\bibitem{Lew}C.H.Lewenkopf and V.G.Zelevinsky. Nucl. Phys. {\bf A569}
(1994) 183c.
\bibitem{LBBZ}B.Lauritzen, P.F.Bortignon, R.A.Broglia and
V.G.Zelevinsky. Phys. Rev. Lett. {\bf 74} (1995) 5194.
\bibitem{Fraz}N.Frazier, B.A.Brown and V.Zelevinsky. Phys. Rev. {\bf C54}
(1996) 1665.
\bibitem{McD} A.Mekjian and W.M.McDonald. Nucl. Phys. {\bf A121} (1968) 385.
\bibitem{Lane} A.M.Lane and R.G.Thomas. Rev. Mod. Phys. {\bf 30} (1958) 257.
\bibitem{Smith} F.T.Smith. Phys. Rev. {\bf 118} (1960) 349.
\bibitem{PvB}V.G.Zelevinsky and P. von Brentano. Nucl. Phys. {\bf A529}
(1991) 141.
\bibitem{Chom}Ph.Chomaz. Nucl. Phys. {\bf A569} (1994) 203c.
\bibitem{Phil}Ph.Chomaz and N.Frascaria. Phys. Rep. {\bf 252} (1995) 275.
\bibitem{SRSM}V.V.Sokolov, I.Rotter, D.V.Savin and M.M\"{u}ller.
{\it nucl-th}/9704036,9704038, to appear in PRC (July 1997).
\end{thebibliography}
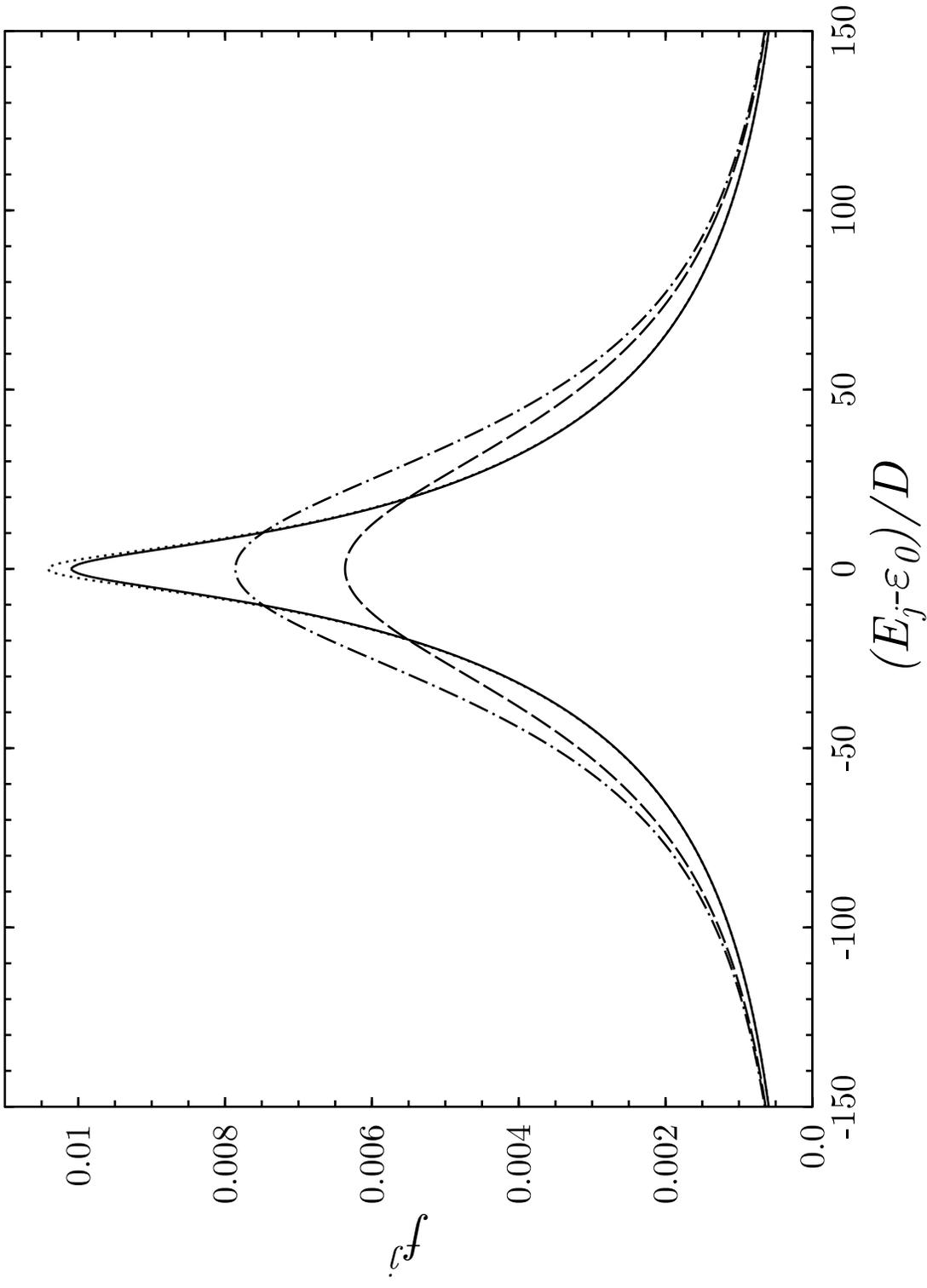
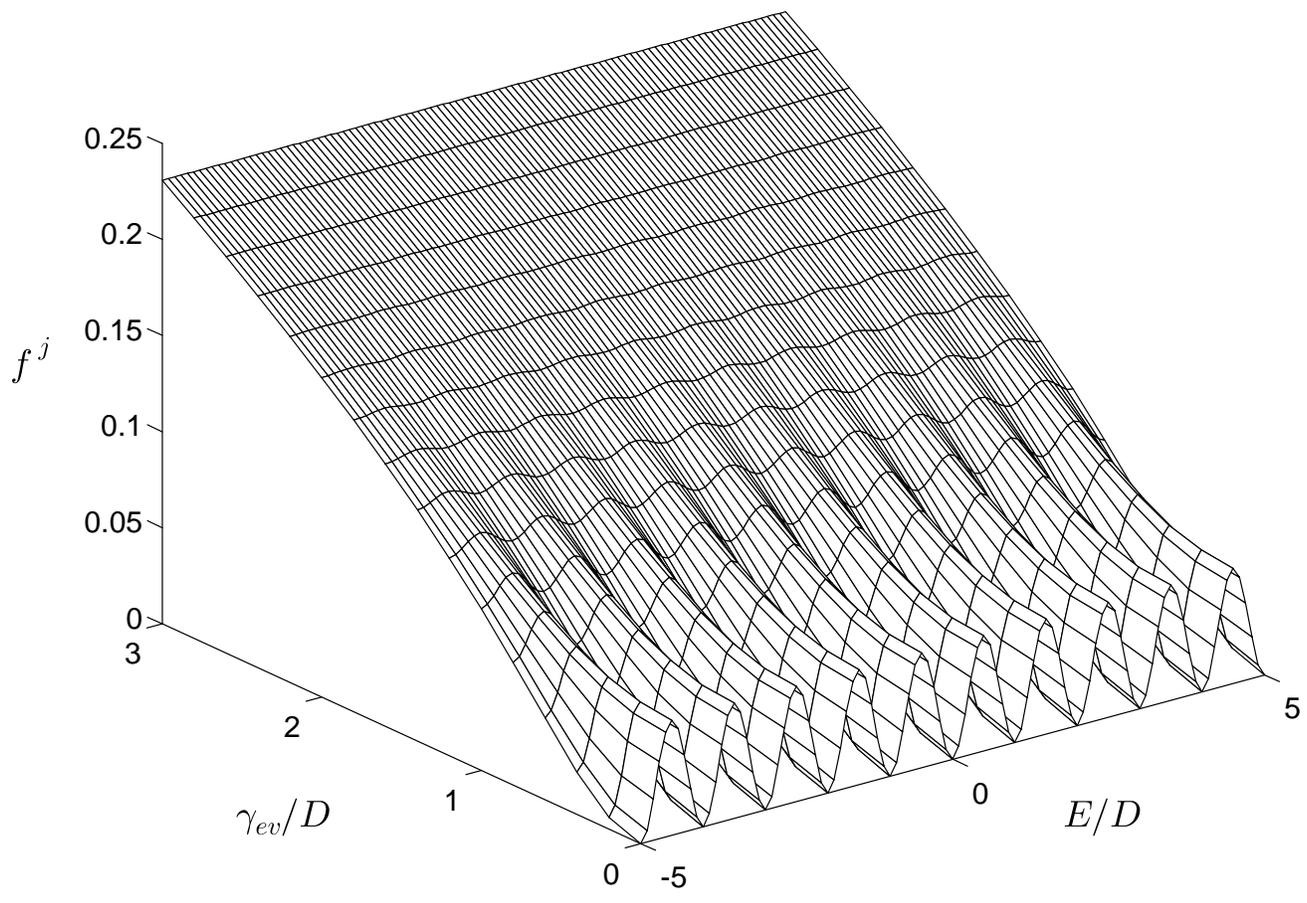
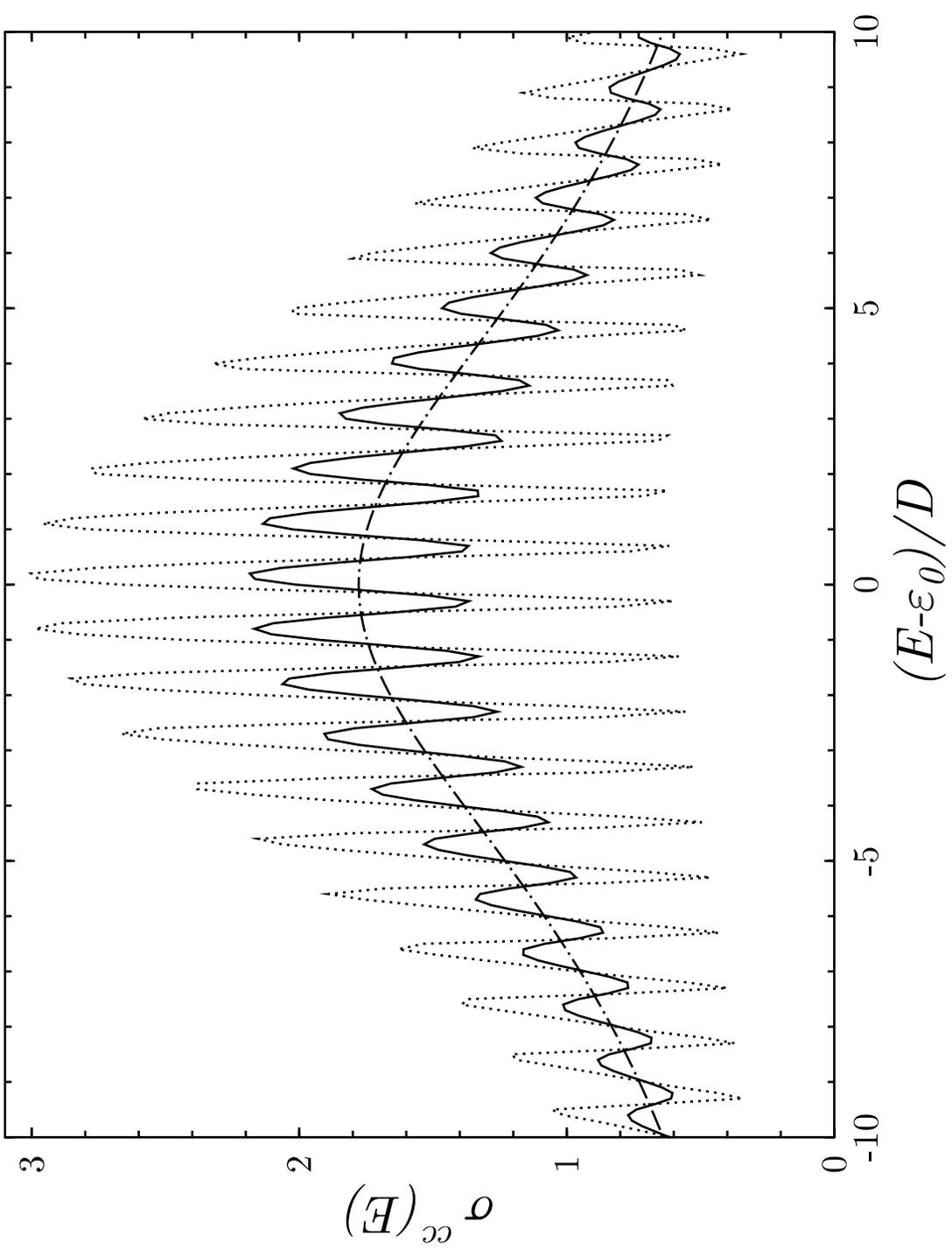
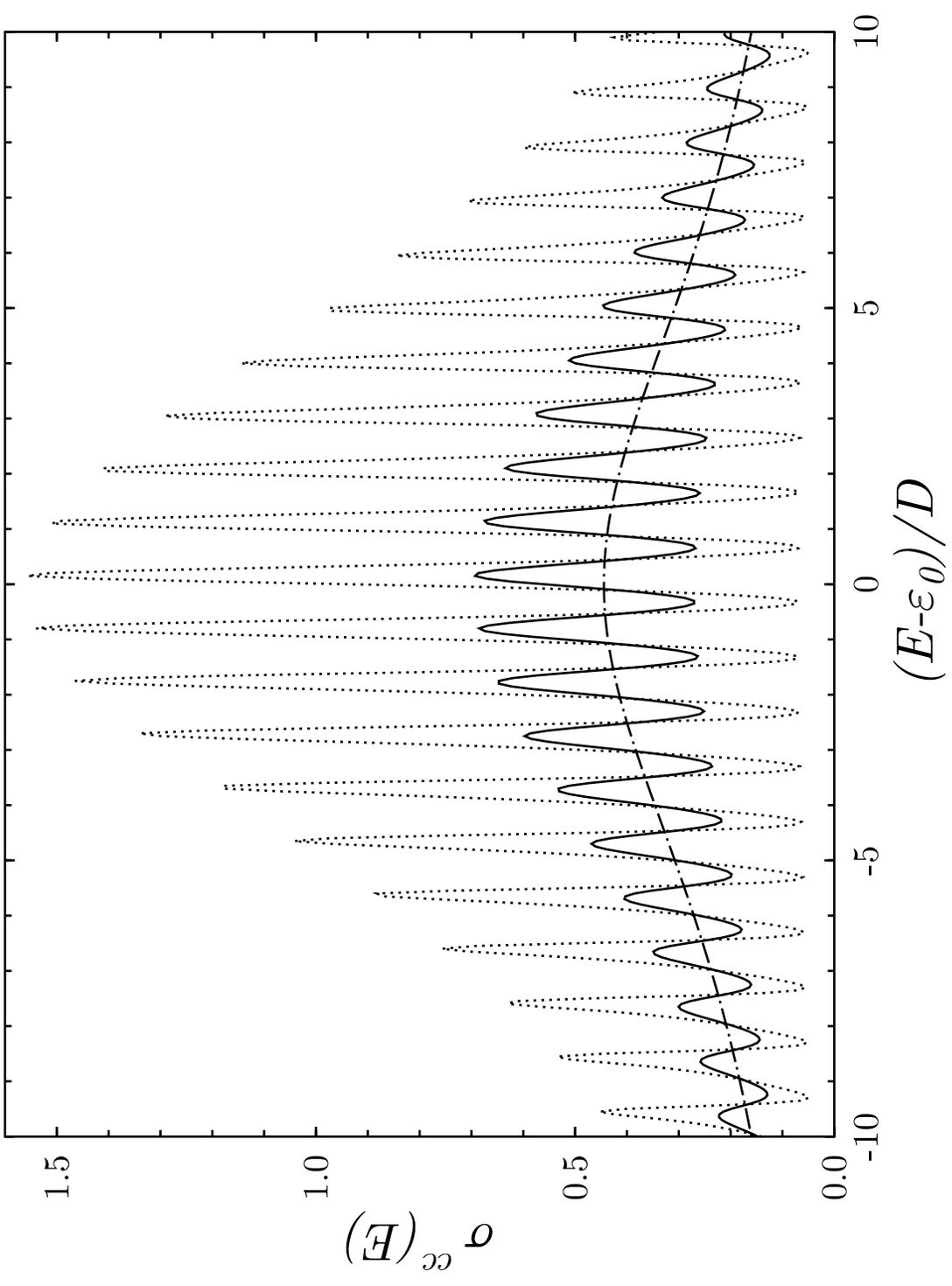
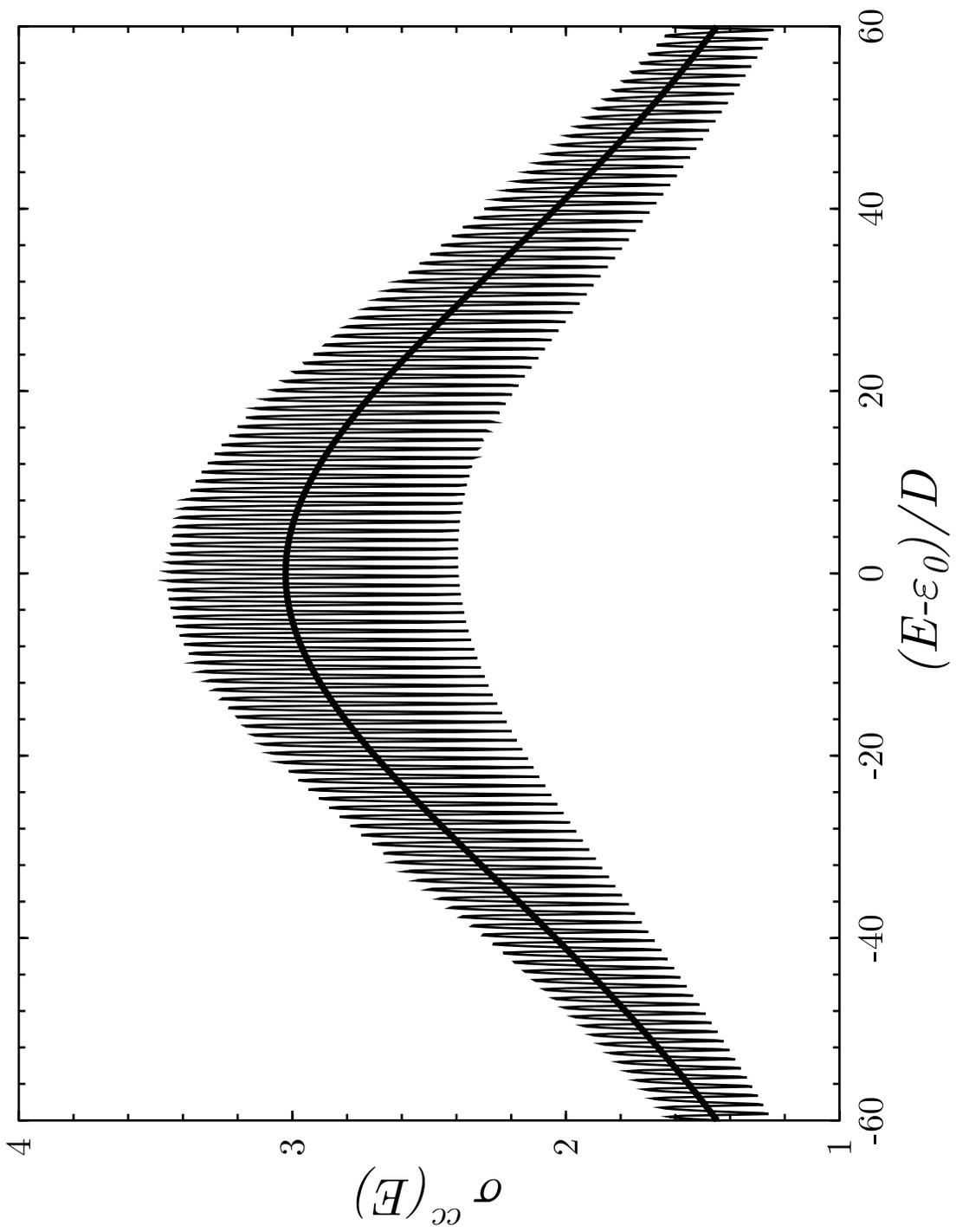
\end{document}